\documentclass{article}
\usepackage{epsfig}

\date{\today}

\begin{document}
\begin{center}
{\Large\bf NEW NONABELIAN SOLUTIONS IN $D=4$, $N=4$ GAUGED SUPERGRAVITY}
\vspace{0.6cm}
\\
Eugen Radu
\\
{\small
\it Albert-Ludwigs-Universit\"at, Fakult\"at f\"ur Physik, Hermann-Herder-Stra\ss e 3
\\
 Freiburg D-79104, Germany}
\end{center}
\begin{abstract}
In this paper we look for static, purely magnetic, nonabelian
solutions with unusual topology  
in the context of $N=4$ Freedman-Schwarz supergravity in
four dimensions.
Two new exact solutions satisfying first order Bogomol'nyi equations are discussed.
The main characteristics of the general solutions are presented and  
differences  with respect to the  spherically symmetric case are studied.
We argue that all solutions present naked singularities.
\end{abstract}
\vspace{1.cm}
\textbf{Introduction.--~}
Looking for solutions of $N=4,~D=4$ gauged $SU(2)\times SU(2)$ supergravity has 
been a subject of long standing interest.
This model,
which we will refer to as the Freedman-Schwarz (FS) model, possesses
a dilaton potential which, apart from being unbounded from below,
has no critical points and hence no obvious ground state \cite{Freedman:1978ra}. 

The action of the FS model
includes a vierbein $e_{\mu }^{m}$, four Majorana spin-3/2 fields 
$\psi _{\mu }^{\rm{I}}$, vector
and pseudovector non-Abelian gauge fields $A_{\mu }^{a}$ and $B_{\mu }^{a}$
with independent gauge coupling constants $g_{A}$ and $g_{B}$, respectively,
four Majorana spin-1/2 fields $\chi ^{\rm{I}}$, the axion
and the dilaton $\Phi$ \cite{Freedman:1978ra}. 
Also, it was shown recently that the FS model can be obtained 
by compactifying $N=1$ ten dimensional supergravity
on the $SU(2)\times SU(2)$ group manifold \cite{Cowdall:1997fn, Chamseddine:1998mc}
(previously also, a Kaluza-Klein interpretation was given in \cite{Antoniadis:1990mn}).

This theory (without fermionic matter) presents an interesting and 
relatively simple model to study various bosonic solutions
with unbroken supersymmetry.
In a pioneering paper, a stable electrovac state
was found by Freedman and Gibbons \cite{Freedman:xa},
which is a product manifold
$AdS_2\times R^2$, and preserves one quarter or one half
of the supersymmetries, the latter case occurring if one of the two
gauge coupling constants vanishes.
There are also other supersymmetric vacua of the Freedman-Schwarz model,
in particular the domain wall solution \cite{Cowdall:1998bu,Singh:1998qd} 
preserving also one half of the supersymmetries. 
This solution has vanishing gauge fields and is purely dilatonic.
Furthermore, BPS configurations involving a non-zero axion were
found by Singh \cite{Singh:1998vf,Singh:1998qd}.

In recent years we witnessed a rapid growth of interest in this topic
following the discovery by Chamseddine and Volkov of a
non-abelian magnetic BPS solution \cite{Chamseddine:1998mc,Chamseddine:1997nm}.
This is one of the few analytically known 
configurations involving both non-abelian gauge fields and gravity
(for a general review of such solutions see \cite{Volkov:1999cc,Gal'tsov:2001tx}).
The Chamseddine-Volkov solution is globally regular, 
preserves $1/4$ of the initial supersymmetry of
the FS model and has unit magnetic charge.
This configuration
is neither asymptotically AdS nor asymptotically flat,
a common situation in the presence of a dilaton potential 
(see $e.g.$ \cite{Chan:1995fr}-\cite{Cai:1998ii}).
Its ten-dimensional lift was shown 
to represent 5-branes wrapped on a shrinking $S^2$ \cite{Chamseddine:1998mc}.
As conjectured by Maldacena and Nu\~nez, this solution provides 
a holographic description for $N=1,~D=4$ super-Yang-Mills theory \cite{Maldacena:2001yy}.
A detailed study of spherically symmetric solutions in the FS model, 
including black holes has been performed
in a recent paper by Gubser, Tseytlin and Volkov \cite{Gubser:2001eg}.

Our work is motivated by the observation that the dilaton potential
present in the FS action can be viewed as an effective negative, position-dependent 
cosmological term.
However, an Einstein-Yang-Mills (EYM) theory 
with negative cosmological constant $\Lambda$
presents
black hole solutions  with nonspherical 
event horizon topology \cite{VanderBij:2001ia}.
They generalize the known spherically symmetric solutions 
\cite{Winstanley:1999sn,Bjoraker:2000qd} replacing the round 
two-sphere by a two-dimensional space $\Sigma$ 
of negative or vanishing curvature.
Also, the study of topological black holes is connected with the AdS/CFT correspondence 
and has been seminal to
recent developments in black hole physics 
(see $e.g.$ \cite{Cai:1996eg}-\cite{Aros:2001ij}).
Therefore it is natural to seek similar solutions in a EYM-dilaton theory
with an effective cosmological term. 

Abelian- BPS black hole solutions of the FS model with toroidal event horizon 
were constructed in \cite{Klemm:1998in}.
In general, the toroidal BPS states represent naked singularities,
but there is also a supersymmetric black hole with vanishing
Hawking temperature.
This motivated us to look for new configurations in
gauged $SU(2)\times SU(2)$ supergravity with a  
nonspherical topology of the constant $(r,~t)$ surfaces. 
Naively, we may expect these solutions to share 
common properties with the EYM$-\Lambda$ configurations.

\textbf{General framework.--~}
One can consistently truncate the FS model by requiring that $g_{B}=B_{\nu }^{a}=0$,
while $A_{\mu }^{a}$  is purely magnetic, 
in which case the axion can be set to zero too.
After a suitable rescaling of the fields, the bosonic part of the 
action reads (we follow the conventions and notations 
used by Gubser, Tseytlin and Volkov in their paper \cite{Gubser:2001eg})
\begin{eqnarray}
\label{actionFS}
S_4=\int d^4x   \sqrt{-g}  \Big (\frac{1}{4} R 
-\frac{1}{2}\partial_\mu\Phi \,\partial^\mu\Phi
-\frac{1}{8}\,{\rm e}^{2\Phi}
F^{a}_{\mu\nu} F^{a \mu\nu}
+\frac{1}{4} {\rm e}^{-2\Phi}),
\end{eqnarray}
where
$F^{a}_{\ \mu\nu}=
\partial_{\mu}A^{a}_{\nu}-\partial_{\nu}A^{a}_{\mu}
+\varepsilon_{abc}A^{b}_{\mu}A^{c}_{\nu}$.

We consider spacetimes whose metric can be written locally in the form
\begin{eqnarray}
\label{metric}
ds^{2}=e^{2\Phi}\left(\frac{e^{2\lambda}}{\nu}dr^2+R^{2}d \Omega_k^2-
\nu dt^{2} \right),
\end{eqnarray}
where $d \Omega_k^2=d\theta^{2}+f^{2}(\theta) d\varphi^{2}$
is the metric on a two-dimensional surface $\Sigma$ of constant curvature $2k$.
Here $\Phi$, $\nu$, $\lambda$, $R$ are functions of the coordinate
$r$ (we do not fix the gauge at this stage).
The discrete parameter $k$ takes the values $1, 0$ and $-1$ 
and implies the form of the function $f(\theta)$
\begin{eqnarray}
f(\theta)=\left \{
\begin{array}{ll}
\sin\theta, & {\rm for}\ \ k=1 \\
\theta , & {\rm for}\ \ k=0 \\
\sinh \theta, & {\rm for}\ \ k=-1.
\end{array} \right.
\end{eqnarray}
In this solution, the topology of the two-dimensional space $t=$const. and $r=$const.
depends on the value of $k$.
When $k=1$, the metric takes on the familiar spherically symmetric form,
and the $(\theta, \varphi)$ sector has constant positive curvature. 
For $k=0$, the $\Sigma$ is a flat surface and
the $(\theta, \varphi)$ sector is a space with constant negative curvature, 
also known as a hyperbolic plane. 
The solutions have four Killing vectors, one timelike and three spacelike,
indicating the symmetries of spacetime.

The most general expression for the appropriate $SU(2)$ connection 
is obtained by using the standard rule 
for calculating the gauge potentials for any spacetime group 
\cite{Forgacs:1980zs,Bergmann}.
Taking into account the symmetries of the line element (\ref{metric}) we find 
\begin{eqnarray} \label{A}
A=\frac{1}{2} \left\{ u(r,t) \tau_3 dt+ \chi(r,t) \tau_3 dr+
\left( \omega(r,t) \tau_1 +\tilde{\omega}(r,t) \tau_2\right) d \theta
+\left(\frac{d \ln f}{d \theta} \tau_3
+ \omega(r,t) \tau_2-\tilde{\omega}(r,t)\tau_1  \right) f d \varphi \right \}, 
\end{eqnarray}
 
where $\tau_a$ are the Pauli spin matrices.
For purely magnetic, static configurations ($i.e.$ $u=0$) 
it is convenient to take the $\chi=0$ gauge
and eliminate $\tilde{\omega}$ by using a residual gauge freedom. 
The remaining function $\omega$ depends only on the coordinate $r$.
As a result, we obtain the YM curvature $F$
\begin{eqnarray} 
F=\frac{1}{2}\left (
\omega' \tau_1 dr\wedge d\theta +
f \omega' \tau_2 dr\wedge d\varphi +
(w^2-k)f \tau_3 d\theta \wedge d\varphi \right ),
\end{eqnarray}
where a prime denotes a derivative with respect to $r$.

To find BPS solutions, it is convenient to 
introduce new variables
\begin{eqnarray}
\lambda=\frac{1}{2}\log{\nu}+2s+l,~~~~~R=e^g,~~~~~\Phi=s-g-\frac{1}{4}\log{\nu}.
\end{eqnarray}
Inserting this ansatz into the
action (\ref{actionFS}), integrating and dropping the surface term,
we find that the equations of motion can be derived 
from an effective action whose Lagrangian is given by
\begin{eqnarray}          
\label{lagr}
L={\rm e}^{-l}\left(s'^2-\frac12\,{\rm e}^{-2g}w'^2
-\frac{1}{2}\,g'^2\right)
-\frac{1}{4}\,{\rm e}^{4s+l}
\left[{\rm e}^{-4g}(w^2-k)^2
-2k{\rm e}^{-2g}-1\right] - \frac{1}{4}\left(\frac{\nu'}{\nu}\right)^2 e^{-l}.
\end{eqnarray}
We remark that (\ref{lagr}) allows for the reparametrization
$r \to \tilde{r}(r)$ which is unbroken by our ansatz.
A first integral of motion is given by the equation of $\nu$
\begin{eqnarray}      
(\frac{\nu'}{\nu}{\rm e}^{-l})'=0,
\end{eqnarray}
which implies
\begin{eqnarray}            
\log \nu=\log \nu_0+\alpha \int e^l dr,
\end{eqnarray}
where $\log{\nu}_0$ and $\alpha$ are integration constants.
In the "extremal" case $\alpha=0$ we can set $\nu=1$, without loss of generality.
 
\textbf{New BPS solutions.--~}
The Lagrangian (\ref{lagr}) 
can be written in the form 
\begin{eqnarray}                   
\label{lag}
L=G_{ik}(y){ dy^i\over  dr} { dy^k\over  dr} - U(y),
\end{eqnarray}
where $y^i= (s,w,g)$ and
$G_{ik}=e^{-l} {\rm diag}(1,-\frac12{\rm e}^{-2g},
- \frac{1}{2})$.
For $\alpha=0$, the potential $U$ can be  represented as
\begin{eqnarray}  
U=-G^{ik}\frac{\partial W}{\partial y^i}\frac{\partial W}{\partial y^k},
\end{eqnarray}
where the superpotential $W$  has the expression
\begin{eqnarray} 
W=\frac14{\rm e}^{2s}\sqrt{
{\rm e}^{-4g}(w^2-k)^2 + 2{\rm e}^{-2g}\,(w^2+k)+1}.
\end{eqnarray}  
As a result,  we find
the first order Bogomol'nyi equations 
\begin{eqnarray}
\label{BPS;eq1}
s'&=&\frac{1}{2}e^{l+2s}\sqrt{e^{-4g}(w^2-k)^2 + 2e^{-2g}(w^2+k)+1},
\\
\label{BPS;eq2}
\omega'&=&-\frac{e^{l+2s-2g}\omega \left(\omega^2-k+e^{2g}\right)}
{\sqrt{e^{-4g}(w^2-k)^2 + 2e^{-2g}(w^2+k)+1}},
\\
\label{BPS;eq3}
g'&=&\frac{e^{l+2s} \left(e^{-4g}(\omega^2-k)^2+e^{-2g}(w^2+k)\right)}
{\sqrt{e^{-4g}(w^2-k)^2 + 2e^{-2g}(w^2+k)+1}},
\end{eqnarray}
which solve also the second-order system.
By using the gauge $l=-2s$, 
it is possible to find the
general solutions of the first-order equations 
(\ref{BPS;eq1})-(\ref{BPS;eq3}) for any value of $k$.

For vanishing curvature of the two-dimensional space $\Sigma$ 
($k=0$), the superpotential $W$ has the simple form
$W=\frac{1}{4}e^{2s-2g}(w^2+e^{2g})$
and the solution is
\begin{equation}
\label{solk=0;1}
ds^{2}=\frac{e^{r+2\Phi_0}}{\sqrt{c-e^{-2(r+r_0)}}}
\left (dr^{2}+e^{2g(r)}d \Omega_0^2-dt^2\right) ,  
\end{equation}
\begin{eqnarray}
\label{solk=0;2}
w(r)=e^{-(r+r_0)}, \ \ \ \ \
e^{2(\Phi(r)-\Phi_0)}=e^{-g(r)}e^{r},\ \ \
R^2(r)=e^{2g(r)}=c-\omega^2(r).
\end{eqnarray}
The solutions for $k=\pm 1$ are more complicated. 
Taking the ratio of $g$ and $w$ equations and using the new variables
$\omega^2=u$ and $e^{2g}=v$
we obtain the first-order equation
\begin{eqnarray}
\label{eqgen}
u(u+v-k)\frac{dv}{du}+v(u+k)+(u-k)^2=0.
\end{eqnarray}
By using the substitution ($u=\rho^2 e^{\xi},~~v=-u-k(\rho \xi'+1)$),
eq.(\ref{eqgen}) reduces to
\begin{eqnarray}
\xi''=2ke^{\xi},
\end{eqnarray}
which can be solved.
As a result, for $k=1$ the exact solution is \cite{Gubser:2001eg}
\begin{equation}
\label{solk=1;1}
ds^{2}= e^{-g(r)+2\Phi_0}\sinh (r+r_0+c)\left 
(dr^{2}+e^{2g(r)}d \Omega_1^2-dt^2\right) ,  
\end{equation}
\begin{equation}
w(r)=\frac{r+r_0}{\sinh(r+r_0+c)},\ \ \ \
e^{2(\Phi(r)-\Phi_0) }=e^{-g(r)}\sinh (r+r_0+c),
\label{solk=1;2}
\end{equation}
\begin{equation}
R^2(r)=e^{2g(r)}=2(r+r_0)\coth(r+r_0+c)-\omega^2(r)-1.
\label{solk=1;3}
\end{equation}
The solution for $k=-1$ has the form
\begin{equation}
\label{solk=-1;1}
ds^{2}= \frac{e^{-g(r)+2\Phi_0}}{\cosh(r+r_0+c)}\left 
(dr^{2}+e^{2g(r)}d \Omega_{-1}^2-dt^2\right) ,  
\end{equation}
\begin{equation}
w(r)=\frac{r+r_0}{\cosh(r+r_0+c)},\ \ \ \
e^{2(\Phi(r)-\Phi_0) }=\frac{e^{-g(r)}}{\cosh (r+r_0+c)},
\label{solk=-1;2}
\end{equation}
\begin{equation}
R^2(r)=e^{2g(r)}=-2(r+r_0)\tanh(r+r_0+c)-\omega^2(r)+1.
\label{solk=-1;3}
\end{equation}
In the above relations $r_0$, $\Phi_0$ and $c$ are integration constants.
Different choices of $\Phi_0$ correspond
to global rescalings of the
solutions, while $r_0$ can be absorbed  by  shifting
 $r\to r-r_0$.
 
For a given $c$ we prefer to set the constant $r_0$ 
so that the point $r=0$ corresponds to origin, where $R(r)$ vanishes.
To better understand the properties of the $k=0$ solution, we can express it 
by using a new coordinate $\rho=\sqrt[4]{e^{2r}-1}$
\begin{equation}
\label{solk=01;1}
ds^{2}=e^{2\Phi_0}
\left (\frac{4\rho^4}{\rho^2+1}d\rho^{2}+\rho^2d \Omega_0^2-\frac{\rho^4+1}{\rho^2}dt^2\right) ,  
\end{equation}
\begin{eqnarray}
\label{solk=01;2}
w(\rho)=\frac{1}{\sqrt{\rho^4+1}}, \ \ \ \ \
e^{2\Phi(\rho)}=\frac{e^{2\Phi_0}(\rho^4+1)}{\rho^2}.
\end{eqnarray}

For $k=1$, setting $c=r_0=0$, we obtain a regular solution.
The spacetime is geodesically complete and globally hyperbolic \cite{Chamseddine:1997nm}.

However, if $k\neq 1$ we find that, for every choice of the integration constants,
the line elements (\ref{solk=0;1}) and (\ref{solk=-1;1}) present naked singularities.
A direct computation reveals that the point $r=0$ is a curvature singularity.
This singularity appear to be repulsive: no timelike geodesic hits it,
though a radial null geodesic can. 
Our solutions violate the criterion of \cite{Maldacena:2001mw} 
because $g_{tt}$ in the Einstein frame is
unbounded at the singularity and thus they cannot accurately
describe the IR dynamics of a dual gauge theory.

Note that for $k=-1$, the coordinate $r$ is restricted to the region $0<r<r_c$ 
(elsewhere $R^2(r)$ will become negative).
As $r\to r_c$, $R(r)\to 0,~\Phi(r)$ diverges and there is a second curvature singularity.

\textbf{Nonextremal solutions.--~}
One may ask whether there are non-BPS solutions 
with a regular origin. This is the point $r=r_0$, where the function $R(r)$ vanishes but all
curvature invariants are bounded (without loss of generality we can set $r_0$=0).

When written in the variables ($\nu,~R,\Phi$) the field equations of the FS model
are (we use the gauge $\lambda=0$)
\begin{eqnarray}                
\label{eq1}
-\frac{k}{\nu}+\frac{(k-\omega^2)^2}{2\nu R^2}-\frac{R^2}{2\nu}
+\frac{R\nu'(R'+R\Phi')}{\nu}+R'^2-\omega'^2
+4RR'\Phi'+2R^2\Phi'^2&=&0,
\\
\label{eq4}
\omega''-\frac{\omega(\omega^2-k)}{\nu R^2}+\frac{\nu'\omega'}{\nu}+2\omega'\Phi'&=&0,
\\
\label{eq2}
R''+\frac{(k-\omega^2)^2}{\nu R^3}-\frac{k}{\nu R}
+R'(\frac{\nu'}{\nu}+2\Phi')+\frac{R'^2}{R}+\frac{\omega'^2}{R}&=&0,
\\
\label{eq3} 
\Phi''-\frac{(k-\omega^2)^2}{\nu R^4}+\frac{k}{\nu R^2}
-\frac{\nu' R'}{\nu R}-\frac{R'^2}{R^2}-\frac{2R'\Phi'}{R}&=&0,
\\ 
\label{eq5}
\nu'-\frac{2\alpha e^{-2\Phi}}{R^2}&=&0.
\end{eqnarray}

For the spherically symmetric case, there is a continuum of regular solutions 
in terms of the adjustable shooting parameter
that specifies the initial conditions at the origin.
It was proven in \cite{Gubser:2001eg} that the finite energy solutions form a discrete set,
indexed by the number of zeros of the gauge function.

However, a direct inspection of the system (\ref{eq1})-(\ref{eq5}) reveals the absence 
of solutions with a regular origin for $k\neq 1$.
In this case it is not possible to take $R(r_0)=0$ without introducing a curvature 
singularity.
We  observe the similarity with the EYM system with $\Lambda<0$, where the absence 
of regular configurations has been noticed in \cite{VanderBij:2001ia}.
This fact has to be attributed to the particular form of the fourth order 
YM potential in (\ref{lagr}).


Apart from the regular solutions, Gubser, Tseytlin and Volkov  
have found a rich spectrum of 
non-abelian spherically symmetric black hole solutions \cite{Gubser:2001eg}, 
corresponding 
to a non-zero parameter $\alpha$ in (\ref{lagr}) ($i.e.$ a non-constant $\nu(r)$).

A natural way to deal with singularities is to hide them inside an event horizon.
To implement the black hole interpretation we restrict the parameters 
so that the metric describes the exterior of a black hole with a non-degenerate horizon.
That implies the existence
of a point $r=r_h$ where $\nu$ vanishes, while all other functions
are finite and differentiable.
Without loss of generality we can set $r_h=0$.

Unlike the spherically symmetric case,
we find only nodeless solutions. 
This can be analytically proven by 
integrating the equation for $\omega$, 
$(e^{2\Phi}\nu\omega')'=\omega(\omega^2-k)/R^2$ between $r_h$ and $r$; 
thus we obtain $\omega'>0$ for every $r>r_h$.
For $k=1$, 
both nodeless solutions and solutions where $\omega$ crosses the axis can exist.
We notice again the analogy with the EYM-$\Lambda$ system, where a similar behavior 
was found \cite{VanderBij:2001ia}. 


The field equations give the following expansion near the event horizon
\begin{eqnarray} 
\label{expansion}
R(r)&=&R_h+\frac{e^{2\Phi_h}}{2R_h\alpha}\left(kR_h^2
-(\omega_h^2-k)^2\right)r+O(r^2),\label{h1}
\\
\omega(r)&=&w_h+\frac{e^{2\Phi_h}w_h}{2\alpha}(\omega_h^2-k)r+O(r^2),\label{h2}
\\
\Phi(r)&=&\Phi_h
+\frac{e^{2\Phi_h}}{4R_h^2\alpha}\left(R_h^4+(\omega_h^2-k)^2\right)r+O(r^2),\label{h3}
\\
\nu(r)&=& \frac{ 2\alpha e^{-2\Phi_h}}{R_h^2} r+O(r^2).\label{h4}
\end{eqnarray}
The solutions present three free parameters: 
the value of the dilaton at the horizon $\Phi_h$,
the event horizon radius $R_h$ and the value of the gauge potential at the horizon
$\omega_h$.
For every value of $k$, one can set $\alpha=1/2$ without loss of generality, 
since this value can be obtained by a global rescaling of the line element (\ref{metric}).

Using the initial conditions on the event horizon
(\ref{h1})-(\ref{h3}), the equations (\ref{eq1})-(\ref{eq4}) were integrated 
for a range of values of $\Phi_h,~R_h$ 
and varying $\omega _{h}$.
Since the equations (\ref{eq1})-(\ref{eq4}) are invariant under the
transformation $\omega \rightarrow - \omega $, only values of
$\omega _{h}>0$ are considered.
The numerical analysis shows the existence of a continuum of solutions
for every value of ($k,R_h, \omega_h, \Phi_h)$.
Also, for every choice of $\Phi_h$ and a given ($k,R_h,\omega_h)$, we find qualitatively
similar solutions
(different values of $\Phi_h$ lead 
to global rescalings of the solutions).

In the spherically symmetric case \cite{Gubser:2001eg}, 
the black hole properties depend on the two essential
parameters ($R_h, \omega_h)$. 
For $R_h^2+\omega_h^2>1$, $R(r)\to\infty$ as $r\to\infty$, 
all curvature invariants vanish in the same limit 
and the asymptotics is somewhat similar 
to the exact solution (\ref{solk=1;1})-(\ref{solk=1;3}).
For $R_h^2+\omega_h^2<1$, as numerically found by Gubser, Tseytlin and Volkov
\cite{Gubser:2001eg}, the asymptotic is different:
$R(r)$ vanishes at some $r=r^\ast$, where there is a curvature singularity.

The situation for a nonspherically symmetric event horizon resembles 
this last case.
For each set  ($R_h,~w_h$) we find
a solution living in the interval $r\in[0,r^\ast]$, where
$r^\ast$ has a finite value.
For fixed $R_h,~\Phi_h$,
the value of $r^\ast$
decreases when increasing $\omega _{h}$.
Typical solutions are presented in Figure 1.

The function $R(r)$ is no longer unbounded, but vanishes at $r^\ast$ 
where there is a curvature singularity.
This fact can be better understood by studying the combined $R$ and $\omega$ equations
\begin{eqnarray}
\label{rel-w-R}
(e^{2\Phi}\nu (\omega^2+R^2)')'=2k\frac{e^{2\Phi}}{R^2}(\omega^2+R^2-k).
\end{eqnarray}
\newpage
\setlength{\unitlength}{1cm}

\begin{picture}(8,8)
\centering
\put(2,0){\epsfig{file=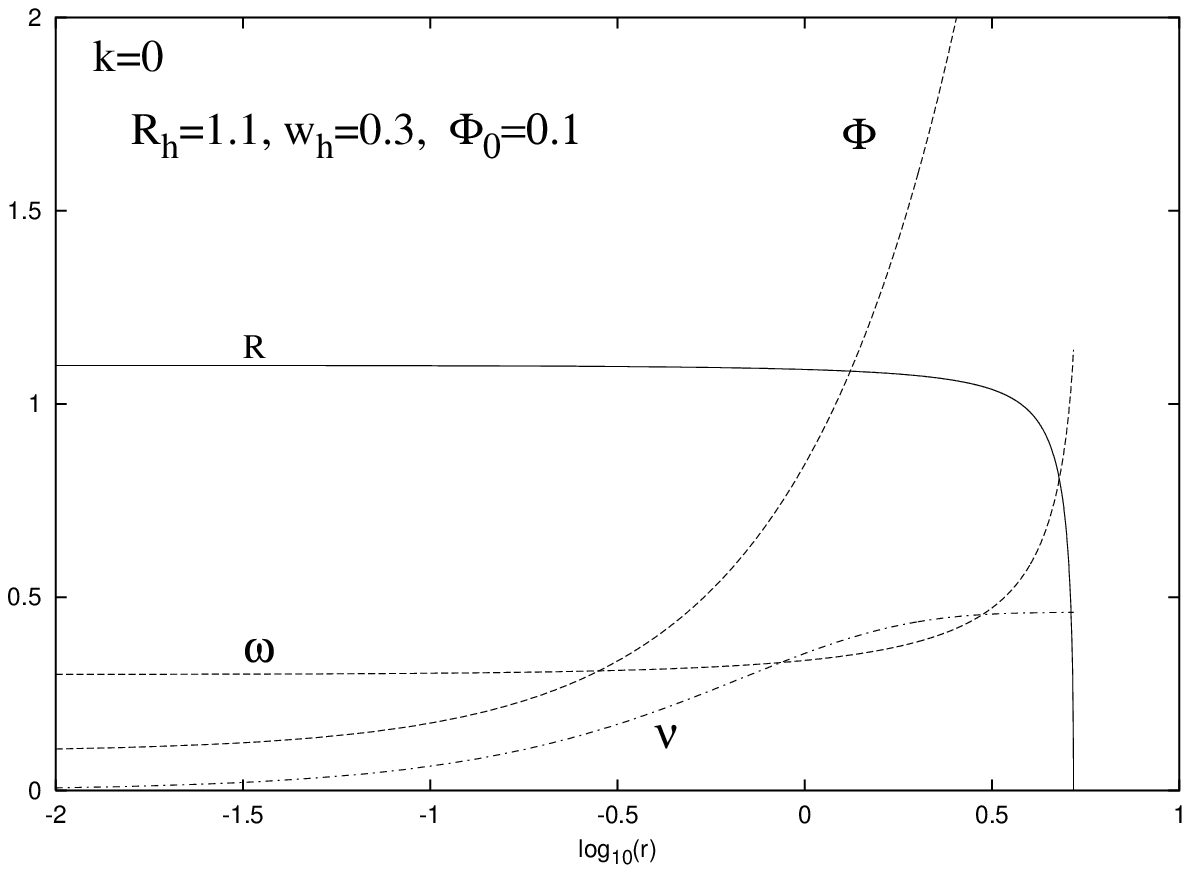,width=12cm}}
\end{picture}
\begin{picture}(19,8.5)
\centering
\put(2.5,0){\epsfig{file=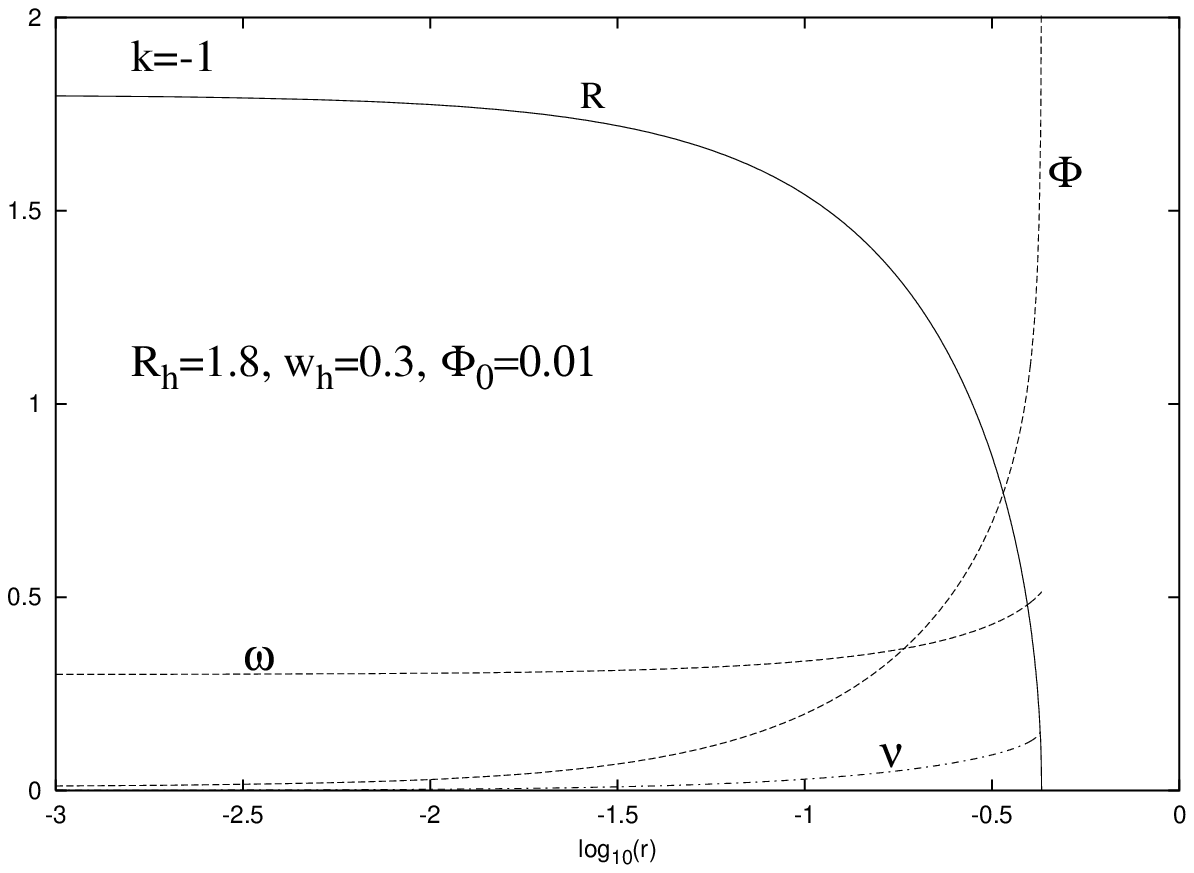,width=12cm}}
\end{picture}
\begin{center}
Figure 1. The gauge function $\omega$, the dilaton $\Phi$
and the metric functions $R,~\nu$ are shown as function of the coordinate
$r$ for generical solutions with  $k=0$ and $k=-1$.
\end{center}
\vspace{0.7cm}

It is evident from this equation that $R \to \infty$ 
is not allowed unless $k=1$ and $R_h^2+\omega_h^2>1$.
For $k=0$ we find $\omega^2+R^2=const.$, while  $k=-1$ configurations
satisfy always
$\omega^2+R^2<w_h^2+R_h^2$.
The relation (\ref{rel-w-R}) clarifies also the 
observed behavior in the spherically symmetric case,
both in the regular and black hole cases.

\textbf{Concluding remarks.--~}
As was shown in \cite{Chamseddine:1998mc}, the FS model
can be obtained via dimensional reduction
of the  $D=10$  supergravity, which contains apart from gravity 
and the dilaton field a Ramond-Ramond 3-form $F_3$ (the NS-NS 3-form $H_3$ is set to zero).
As a result, any
on-shell configuration in the model (\ref{actionFS})
$(g_{\mu\nu},A^{a}_{\mu}, \Phi)$,
can be uplifted to become a solution
of ten-dimensional equations of motion for the $D=10$ supergravity.
Using the uplifting formulas in \cite{Gubser:2001eg}, 
we can extract the $D=10$ metric, the dilaton and the R-R 3-form.
 In the Einstein frame, 
the ten dimensional solution reads
\begin{eqnarray}
\label{metric10d}
ds_{10}^2=e^{\hat{\Phi}/2}\left(e^{-2\hat{\Phi}}g_{\mu \nu}dx^{\mu}dx^{\nu}
+dx^ndx^n+\Theta^a\Theta^a
\right),
\end{eqnarray}
where $g_{\mu \nu}dx^{\mu}dx^{\nu}$ is given by (\ref{metric}), ($a,b,c,n=1,2,3$),
\begin{eqnarray}
\nonumber
A^{a}=A^{a}_{\mu} dx^{\mu},
~~~~~F^{a}=\frac12F^{a}_{\mu\nu}dx^\mu\wedge dx^\nu,
~~~~~\Theta^{a}\equiv \epsilon^a-A^{a},
\end{eqnarray}
 while $\epsilon^a$
are the invariant 1-forms on $S^3$
\begin{eqnarray}
\epsilon_1=\cos\psi d\theta_1+\sin\psi\sin\theta_1 d\phi_1,\
\epsilon_2=-\sin\psi d\theta_1+\cos\psi\sin\theta_1 d\phi_1,\
\epsilon_3=d\psi+\cos\theta_1 d\phi_1,
\end{eqnarray}
$\psi,\theta_1,\phi_1$ being the Euler angles on the three sphere. 
The $D=10$ dilaton field is $\hat{\Phi}=\Phi+\ln 4$,
while the 3-form $F_3$
is given by
\begin{eqnarray}
F_3=\Theta^{1}\wedge\Theta^{2}\wedge\Theta^{3}
- \Theta^{a}\wedge F^{a}.
\end{eqnarray}

The ten-dimensional metric (\ref{metric10d}) has a rather complicated form.
We note that the four-dimensional nonabelian field
gives rise to off-diagonal components of the metric.
The discussed configurations are 
3-brane-type solutions with 1+3 ``parallel" directions 
($t,{\rm x}^n$) and 6 transverse directions
$(r,\theta,\phi,\psi,\theta_1,\phi_1)$.
Using the rules of \cite{Chamseddine:1998mc},
one can further lift the solutions to eleven dimensions 
to regard them in the context of M-theory.


To summarize, we have added two more members to the family of known 
supersymmetric exact solution 
with gravitating nonabelian fields.
Our solutions can be regarded as complements of the spherically symmetric
configurations discussed in \cite{Chamseddine:1998mc,Gubser:2001eg}.
However,
a nonspherical topology of the surface $\Sigma$ will change drastically 
the structure and properties of the solutions.

Given the presence of the naked singularities, 
the physical significance of these solutions is not obvious.
It is the dilaton field potential which accounts for the presence of pathologies
in these solutions.
Although there exist some common properties,
a different asymptotic behavior was found 
in the EYM system with a negative cosmological constant.
We can hope that a more general matter content
will lead to a desingularization of the nonspherically symmetric solutions.

It can also be proven that, as expected,
the new BPS solutions preserve $N=1,~D=4$ supersymmetry.
The computation of the Killing spinors and further details on these
new  solutions will be presented elsewhere.
\\
\\
{\bf Acknowledgments}

The author is grateful to J.J. van der Bij 
for useful discussions and P. Cowdall for pointing out Ref. \cite{Cowdall:1997fn}.
\\
This work was performed in the context of the
Graduiertenkolleg of the Deutsche Forschungsgemeinschaft (DFG):
Nichtlineare Differentialgleichungen: Modellierung,Theorie, Numerik, Visualisierung.



\end{document}